\documentclass[final,3p,times]{elsarticle}

\usepackage{tabulary}
\usepackage{todonotes}
\usepackage{listings}
\usepackage{url}

\newcounter{RQn}
\newcommand{\RQset}[2] {%
  \refstepcounter{RQn}%
  \label{RQ:#1}%
  \noindent{\bf RQ\theRQn:} #2%
}

\newcommand{\RQref}[1] {%
  RQ\ref{RQ:#1}%
}

\newcommand{\maxTableWidth} {%
  0.9\textwidth
}

\begin{document}

\title{Towards Base Rates in Software Analytics\tnoteref{pp}\\
\normalsize{Early results and challenges from studying Ohloh}}

\tnotetext[t1]{This document is a pre-print version of the article that has been accepted for publication in the Science of Computer Programming special issue "the Future of Understand Software," honoring Paul Klint's 65th birthday. Expected publication date is early 2014. Barring typesetting, formatting differences and a personal note to Paul, this pre-print is identical to the final version.}

\author{Magiel Bruntink}

\address{System and Network Engineering research group,\\
Faculty of Science,
University of Amsterdam
}

\ead{M.Bruntink@uva.nl}

\begin{abstract}
Nowadays a vast and growing body of open source software (OSS) project data is publicly available on the internet. Despite this public body of project data, the field of software analytics has not yet settled on a solid quantitative base for basic properties such as code size, growth, team size, activity, and project failure. What is missing is a quantification of the base rates of such properties, where other fields (such as medicine) commonly rely on base rates for decision making and the evaluation of experimental results. The lack of knowledge in this area impairs both research activities in the field of software analytics and decision making on software projects in general. This paper contributes initial results of our research towards obtaining base rates using the data available at Ohloh (a large-scale index of OSS projects). Zooming in on the venerable `lines of code' metric for code size and growth, we present and discuss summary statistics and identify further research challenges.
\end{abstract}

\date{Today}
\maketitle
\thispagestyle{plain}
\pagestyle{plain}


\section{Introduction}

``To measure is to know'' said Lord Kelvin in 1883. While he made this statement in the context of electric phenomena, his intent appears to have been a general claim that applies to all fields of science\footnote{The full quote is ``I often say that when you can measure what you are speaking about, and express it in numbers, you know something about it; but when you cannot express it in numbers, your knowledge is of a meagre and unsatisfactory kind; it may be the beginning of knowledge, but you have scarcely, in your thoughts, advanced to the stage of science, whatever the matter may be.'' According to Wikipedia, Lord Kelvin made this famous statement in his {\em Lecture on Electrical Units of Measurement}, which was published in Popular Lectures Vol. I, p. 73, May 1883.}. One could hardly deny that the principle of quantified knowledge has been engrained successfully into the natural sciences. However, what could Lord Kelvin's claim mean to us, researchers and practitioners in the field of software engineering?

The use and study of measurement in our field has been prevalent for decades. However, until recent years, public availability of software project data has been limited. In recent years, the trend has been an increasing public availability of large scale data sets. Manifestations of this trend are, among others, data sets such as Ohloh~\cite{Ohloh:Y922N68F}, PROMISE~\cite{Menzies:2012uw}, GHTorrent~\cite{Gousios:2012gi}, and scientific conferences focussed on studying these data sources, such as the IEEE and ACM Working Conference on Mining Software Repositories. Recently, the term Software Analytics~\cite{Zhang:2011uh} was coined to label research in this area.


Despite the availability of data, the software engineering community has not yet settled on a quantitative base of knowledge on basic software project observations, such as code size, growth, team size, activity, and project failure. What is missing is knowledge of the {\em base rates} of such observations. A base rate is a statistical concept commonly used in the field of medicine to indicate the prevalence of a phenomenon (e.g., a disease) in the population-at-large. The base rate, also referred to as {\em prior probability}, is the measure used when no categorical evidence (e.g., age, medical history) is available.

Furthermore, bases rates are needed to judge experimental results. An example within software engineering could consist of a new methodology that indicates (with 100\% recall and 70\% precision) whether a project will fail. Without knowing the base rate of project failure in the population one would likely suffer a base rate fallacy\footnote{Illustrated on Wikipedia by Keith Devlin: \url{http://en.wikipedia.org/wiki/Base_rate}} by thinking the failure chance is 70\% given just a positive test result. Given a project failure base rate of 20\%, the chance of a project failing given a positive test would be only 45\%.


We propose research that leverages the availability of large-scale data sets to document and discuss base rates for software project observations. In this paper, we report on initial results from our study of data provided by Ohloh (Section~\ref{Section:early-results}). Then, we discuss open research challenges in line with our work in Section~\ref{Section:research-challenges}. Section~\ref{Section:related-work} summarises related work. Finally, we conclude in Section~\ref{Section:conclusion}.

\section{Early results from studying Ohloh}
\label{Section:early-results}

\subsection{Research design}
We are currently doing research that aims at contributing base rates to the field of Software Analytics. For this purpose we collected data from the open-source software index Ohloh~\cite{Ohloh:Y922N68F} in July 2013. Generally these data consist of a monthly aggregate of data resulting from analysis (done by Ohloh) of the source code and the version control system(s) of an OSS project. In total, data were collected for 12,360 OSS projects, resulting in a grand total of 793,626 project months that are fit for analysis (some project data had to be excluded due to data quality issues, see \ref{data-validation}). The projects have been selected according to Ohloh's measure of project popularity (number of declared users on Ohloh), essentially starting with the most popular project: `Mozilla Firefox', which has 12,143 users. The last project collected was `ooosvn', which has 1 user. All collected data and tools developed for processing are available publicly~\cite{OhlohAnalytics:UKMxGUq9}.

The reason this research was started initially was to get insight in the empirical behaviour of the lines of code metric. We did not aim to derive base rates directly. Instead we first explore the data set and understand its (statistical) properties and related challenges. Concretely, two questions were raised that we will aim to answer:

\vspace{1em}
\RQset{code-size}{What is the typical code size of an OSS project?}\\
\RQset{code-growth}{What is the typical yearly growth of the code size of an OSS project, both absolutely and relatively?}
\vspace{1em}

Table~\ref{Table:GQM} summarises the setup of our research using the template provided by the Goal-Question-Metric (GQM) framework~\cite{vanSolingenRevision:2002vg}. In essence, we aim to answer our questions by using the data collected from Ohloh to calculate the metrics CS, CGa, CGI for all projects. We use the year 2012 as the final cut-off year in the data set, as the data for 2013 are not yet complete.

\begin{table}[tb]
\begin{center}
\begin{tabulary}{\maxTableWidth}{llL}
\hline
{\bf Goal} 

 & {\bf Purpose}   & Extend the quantitative knowledge base\\

 & {\bf Issue}     & on commonly used metrics of project evolution \\

 & {\bf Object}    & of OSS projects \\

 & {\bf Viewpoint} & from the viewpoint of a researcher or decision maker, in either case external to the projects under study. \\

\hline
  {\bf Questions} & {\bf \RQref{code-size}} & What is the typical code size of an OSS project?  \\
  & {\bf \RQref{code-growth}} & What is the typical yearly growth of the code size of an OSS project, both absolutely and relatively? \\

\hline
  {\bf Metrics} & {\bf CS} & {\bf Code Size}: Measured by lines of code, i.e., lines of source text excluding comments and white space, including all of a project's source text in all programming languages recognised by Ohloh.\\
  & {\bf CGa} & {\bf Absolute Code Growth}: The absolute growth (positive or negative) of CS in a year, per project.\\
  & {\bf CGi} & {\bf Indexed Code Growth}: An index of the code growth in a year, per project. For example, a value of 1.05 represents 5\% code growth since the beginning of the year.\\
\hline
\end{tabulary}
\end{center}
\caption{Summary of the Goal-Question-Metric setup of the research.}
\label{Table:GQM}
\end{table}

\subsection{Data collection and processing}
\label{data-collection-processing}

In short, the data obtained from Ohloh for a single project consist of the following three parts:
\begin{itemize}
\item {\bf Metadata}. Includes, among others, a project description, site URLs, creation and update dates, user count, user supplied tags (keywords), programming language use, and configurations of version control systems\footnote{In July 2013, Ohloh supports the following version control systems: Git, Mercurial, Bazaar, Subversion, and CVS.}.
\item {\bf Size facts}. Includes monthly statistics on size of the projects source code: lines of code (source text excluding comments and white space), lines of comments, blank lines. The line counts used within Ohloh are produced by the tool Ohcount, which is available as an open source project~\cite{OhCount:qTYIGJcT}.
\item {\bf Activity facts}. Includes monthly statistics on changes to the project's source code: added or removed lines of code, comments and blanks, number of commits to the version control systems used by the project, and the number of contributors (users performing the commits).
\end{itemize}

The data collected from Ohloh are limited to a set of tuples that contain the facts listed in Table~\ref{Table:facts-definition}. The Project name, Year, and Month together can be considered as the key uniquely identifying a tuple of facts. The facts derived from the Ohloh data are the results of our own processing. The processing and storage of all data is handled by software developed using the meta-programming language Rascal~\cite{Klint:2009kn}.

\begin{table}[tb]
\begin{center}
\begin{tabulary}{\maxTableWidth}{lL}

{\bf Facts collected from Ohloh} & {\bf Description} \\
\hline
Project name & The name of the project for identification purposes.\\
Year & The calendar year in which the month of this tuple falls.\\
Month & The calendar month for which the contained facts hold.\\
Code size & The count of lines of code at the end of Month.\\

\multicolumn{2}{l}{{\bf Derived facts: monthly code growth}}\\
\hline
Monthly absolute code growth  & Monthly code growth is obtained by calculating the absolute difference in lines of code to the previous month (if any).\\

Monthly indexed code growth  & Monthly code growth is obtained by calculating the relative difference in lines of code to the previous month (if any).\\

\multicolumn{2}{l}{{\bf Derived facts: aggregation by year}}\\ 
\hline
Absolute code growth ({\bf CGa}) & The sum of the values of monthly absolute code growth in Year.\\
 Indexed code growth ({\bf CGi}) & The product of the values of monthly indexed code growth in Year.\\
Code size ({\bf CS}) & The maximum of the monthly code sizes in Year. \\
Project age & The difference in years between Year and the year the project has first been active in the data set. \\

\end{tabulary}
\end{center}
\caption{Definition of the data tuples that hold the project facts. The facts that directly match the metrics defined in the measurement setup (Table~\ref {Table:GQM}) are indicated by their acronyms. }
\label{Table:facts-definition}
\end{table}

\subsection{Data validation}
\label{data-validation}

After data collection, the several steps of data validation were performed to ensure quality of the data. Table~\ref{Table:data-size} gives a numerical overview of the size of the data set before and after validation and processing. At the time of writing, the following issues with the Ohloh data are known:
\begin{enumerate}
\item Projects that are missing either one of the size facts or the activity facts are excluded from further processing. Projects can also have both size and activity facts, but not for the same months. Such data were also excluded from the data set.
\item Projects that use the SVN version control system\footnote{Indicated by either `SvnRepository' or `SvnSyncRepository' in Ohloh's data on the enlistments --or source code locations-- supplied for a project.} are instructed to supply Ohloh with the only those code directories that represent active and unique developments. If instead a project just lists its top-level SVN directory, all project branches and tagged versions are also included in Ohloh's analysis, possibly leading to many duplications and inflated metrics. Such projects were excluded from the data set.
\item In a small number of cases, Ohloh code analyses have produced negative values for code size. These tuples were excluded from the data set.
\end{enumerate}

\begin{table}[b]
\begin{center}
\begin{tabulary}{\maxTableWidth}{lr}
{\bf Projects collected from Ohloh} & 12,360 \\
1. Projects excluded due to missing data & 586 \\ 
2. Projects excluded due to improper SVN configuration & 931\\
\hline
Projects remaining & 10,843 \\
Project months for the remaining projects & 766,282\\ 
3. Project months excluded due to negative code size values & 658\\
\hline
Project months remaining & 765,624 \\
Project years remaining & 73,402 \\
\hline
\hline
{\bf Projects finally remaining after cut-off} & 10,762 \\
{\bf Project months finally remaining after cut-off} & 701,376 \\
{\bf Project years finally remaining after cut-off} & 64,020 \\
\end{tabulary}
\end{center}
\caption{The size of the data set before and after validation steps as described in Section~\ref{data-validation}. This data set was collected from Ohloh in July 2013. The data set was cut-off after the year 2012.}
\label{Table:data-size}
\end{table}

\subsection{Results}

The results are based on the data set collected from Ohloh in July 2013. One can obtain the data set at our GitHub project OhlohAnalytics~\citep{OhlohAnalytics:UKMxGUq9} under the tag `LAFOUS2013-FINAL'. In this section we give a summary of the results, while  detailed plots of the data are available in \ref{Appendix:detailed-results}. The statistical analysis of the data was done using R \cite{RDevelopmentCoreTeam:2008wf}. Table~\ref{Table:results} reports on the results using the following basic summary statistics:
\begin{itemize}
\item Medians to measure central tendency,
\item The Inter Quartile Range (IQR) to measure dispersion\footnote{The IQR represents the ``middle fifty'' percent of the value range and is commonly used to indicate dispersion. It is defined as the difference between the third and first quartiles, and thus includes the median value (by definition the second quartile). In a boxplot, the IQR measures the height of the box.},
\item The number of observations in total,
\item The number and rate of outliers, as determined using the common 1.5 times IQR threshold.
\end{itemize}
Representativeness of our data set was tested using the method proposed by Nagappan et. al. in \citep{nagappan-esecfse-2013}. Our data set scores a 99\% coverage, given their universe of active projects on Ohloh, the dimensions code size and 12 month number of contributors, and the standard configuration of numerical similarity functions. Given this representativeness score we can claim to be reporting on the population of active OSS projects on Ohloh.

\begin{table}[tb]
\begin{center}
\begin{tabulary}{\maxTableWidth}{lrcrrr}
{\bf Metric} & {\bf Median} & {\bf Median project(s) (in year)} & {\bf IQR} & {\bf Observations} & {\bf Outliers (\%)} \\
\hline
{\bf CS} & 27,998.5 & `profile2' and `dayplanner' (both in 2012) & 115,047 & 9,820 & 1,317 (13\%) \\
{\bf CGa} & 1,028 & `remind' (in 2008) and 9 others & 11,640 & 64,020 & 11,640 (18\%) \\
{\bf CGi} & 1.054 & `gnus' and `petsc' (both in 2004) & 0.343 & 64,020 & 9,928 (16\%)\\
\end{tabulary}
\end{center}
\caption{Summary of the results. Boxplots are available in \ref{Appendix:detailed-results}. In total 10 projects share the median value in case of the CGa metric.}
\label{Table:results}
\end{table}

\subsection{Discussion}

The results we present give statistical answers to our research questions, but do they really relate to the notion of `typical' values (in case of the medians)? For the sake of discussion let us assume that they do. In this interpretation, typical values for an OSS project are a code size of 28,000 LOC (in 2012), a yearly code growth of 1,000 LOC or a 5\% code growth rate\footnote{Note that we would not expect the median growth rate, when applied to the median code size in 2012, to equal the median absolute code growth. The code growth medians are based on all years in the data set, not just 2012.}.

The data tell us at least two other things: First, the metric values are highly dispersed and skewed. Looking at the IQR for Code Size, and the boxplots in \ref{Appendix:detailed-results}, we learn that projects are commonly 90 KLOC bigger than the median, or 20 KLOC smaller. Code growth can be 10 KLOC higher per year, or have a higher rate by 29\% (points). In all cases, the medians are far away from the centre of the IQR, indicating skew. Second, there are many outliers, i.e., metric values that are far away from the median. Considering that each of these outliers represent entire project years, we feel that a statistical elimination of these outliers would not be justified.

Looking at the projects (years) that appear as our medians, what do we observe?
\begin{description}
\item{{\bf Medians for CS}}: `profile2' is a fairly stable Java project with in total 18 contributors. Its functionality is a redesigned personal profile tool for the Sakai collaboration environment. `dayplanner' is a one-man project written in Perl offering personal time management functionality. Both projects consist of 28 KLOC in 2012.

\item{{\bf Medians for CGa}}: `remind' in 2008 is a 24 KLOC project, mainly in C, that is being actively developed by one contributor. The project is described as `a sophisticated calendar/alarm program.'

\item{{\bf Medians for CGi}}: `gnus' is a mail and news reader for GNU Emacs. In 2004 it consists of 110 KLOC, mostly Emacs Lisp. In 2004 it is being developed by 5 to 11 monthly contributors. `petsc' is a suite of data structures and routines for solving partial differential equations. In 2004 it consists of 600 KLOC that are written in C, C++ and Python. There is active development going on by 7 to 10 monthly contributors.

\end{description}
A stereotypical OSS project such as Mozilla Firefox is far away from those medians: in 2012 its maximal code size is 8.2 MLOC, growing by 2 MLOC, at a rate of 30\%. A project such as Firefox may intuitively capture the essence of open source, but it does not characterise the apparent long tail of the OSS population. A similar observation was recently made by Nagappan et. al. in \citep{nagappan-esecfse-2013}. The notion of a `typical' OSS project seems still out of reach.

So, do these results bring us closer to useable base rates for code size and growth? On the one hand, they do. Consider the situation that nothing is known about a project (except that it is OSS), and still a number is needed to estimate the project's code size or growth. In other words, what to do when there is no categorical evidence on the project in question? In that case these numbers provide a first approximation of a base rate for code size and growth (for the population of OSS projects).

On the other hand, it is clear that the results give rise to many new questions: What are the underlying causes of the high dispersion and outlier rates? Of course OSS projects are very diverse: Projects with 1 contributor working in Java are now being lumped together with 900 contributor projects that employ over 20 programming languages. Furthermore, there are many different development methodologies in use, different expertise levels, and so on. Would we obtain more precise base rates by categorising projects while taking such factors into account? How many of the reported outliers are noisy (i.e., results of measurement errors) and how many are true data points? How to deal with that problem at this scale? We discuss these challenges in the next section.

\section{Research challenges}
\label{Section:research-challenges}
For our research, this result is only a starting point. In this section we discuss a number of pressing challenges for future research:

\paragraph{Establishing baserates for basic software project properties}
In addition to base rates for code size and growth, base rates for other common properties like the number of commits and contributors, project age, and so on, also need to be documented. Such base rates are within reach given the current availability of data sets. However, opportunities (and challenges) lie in the combination and triangulation of data from multiple data sets.

Ultimately, there is a need for documentation of base rates for more intricate phenomena, like project failure success and failure (as studied by Wiggins et. al.~\citep{Wiggins:2010cf}), productivity factors (as studied by Delorey et. al.~\citep{Delorey:2007ue}), vulnerability to security issues, and so on. As mentioned by Hassan in~\citep{Hassan:2008hd}, the limitations of OSS data sets should be kept in mind for these efforts. To us it seems that researching base rates could help defining the limitations of the data more sharply.

\paragraph{Categorisation}
There is an obvious next step of applying categorisations to the data set to decrease dispersion. In essence the question becomes if we can identify (plausible) subsets of projects for which the base rates would be more precise. Such categorisations could be based on the use of programming language(s), development methodology, team size, age, functionality, or even other characteristics. Sources offering meta data on OSS projects, such as the Ultimate Debian Database~\citep{Nussbaum:2010hj} or Ohloh itself could be leveraged for this purpose.



\paragraph{Data quality: outliers and noise}
A necessary avenue of research consists of dealing with the problem of data quality (i.e., dealing with outliers and noise) at a large scale. The Software Analytics field has not yet settled on a suitable method for handling outliers and noise, leading to results that are harder to compare and replicate~\citep{Shepperd:2011ew}. We propose following the calls to action by Liebchen~\citep{Liebchen:2010ui}, Shepperd~\citep{Shepperd:2011ew} and Hassan~\citep{Hassan:2008hd}. As a community, we need a (semi-) automated method that facilitates both the detection of noise and outliers, and the manual inspection of such by individual researchers and the research community-at-large.

\paragraph{Visualisation} Finding an effective method to visualise the high-dimensional space spanned by software project metrics, in particular when including a time dimension. The applications of such a method are, among others, enriching project meta data and categorisation of projects. Potentially, this work could also contribute to research on data quality by identifying outliers visually. We are looking into the applicability of promising visualisation techniques such as t-SNE~\citep{VanderMaaten:2008tm} to software engineering data sets.

\paragraph{Defining new metrics} Proposing improved metrics for the inclusion in large-scale data sets on OSS projects. The metrics we now have available may turn out to be performing inadequately from a statistical viewpoint, but more importantly, they may turn out not to be useful as predictors or for other purposes. We aim to identify improvement opportunities based on an analysis of the current metrics. Finally, we aspire to help enable the testing of old and new theories in the field, which may require entirely new metrics to become available at scale.

\paragraph{Moving beyond open source} Data sets on open source, as large scale and diverse as they might be, still offer a limited perspective on software engineering. For instance, large enterprise information systems written in COBOL, but also modern technology such as Java or C\#, are under-represented in the OSS ecosystem. It remains an open question what insights of OSS analytics will be portable to the commercial, closed-sourced software ecosystem.

\section{Related work}
\label{Section:related-work}
The (quantitative) study of (open-source) software repositories has been going on for quite some time, leading to a rich body of literature. In the interest of brevity, not all approaches can be mentioned here. Surveys of the field have been provided by Kagdi et. al. in 2007~\citep{Kagdi:2007us} and Hassan in 2008~\citep{Hassan:2008hd}. A recent overview (2011) of the OSS ecosystem is being provided by Androutsellis-Theotokis et. al. in~\citep{AndroutsellisTheotokis:2011bi}. Examples of large software engineering data sets that are publicly available are GHTorrent~\citep{Gousios:2012gi}, PROMISE~\citep{Menzies:2012uw}, FLOSSmetrics~\citep{GonzalezBarahona:2010hd}, but more are available. A recent overview of data sets is given by Rodriguez et. al. in~\citep{Rodriguez:2012ee}.

Software Analytics is a term recently introduced by Zhang et. al.~\cite{Zhang:2011uh} to label research aimed at supporting decision making in software. Our work can be seen as an instance of Software Analytics. Work that is closely related to ours has been done by Herraiz. He studied the statistical properties of data available on SourceForge and the FreeBSD package base~\citep{HerraizTabernero:2008ts}. We see our work as an follow-up in terms of scope and diversity, as by studying Ohloh, we use a larger and more diverse data source (which does not primarily focus on C).

Other researchers are also studying the data offered by Ohloh. Recently, Nagappan et. al.~\citep{nagappan-esecfse-2013} used the Ohloh data set for defining a method to quantify the representativeness of empirical studies (on OSS projects). In 2008, Deshpande and Riehle reported on an investigation into the total size and growth of the OSS ecosystem using Ohloh~\citep{Deshpande:2008ue}. Also using Ohloh, Arafat and Riehle reported on the distribution of commit sizes~\citep{Arafat:2009eh}.

\section{Conclusion}
\label{Section:conclusion}

We have reported first results in researching a large scale public data set representative of the active OSS projects on Ohloh. We showed what typical values exist for code size and growth (measured by LOC), but we also observed large dispersion, skew and outlier rates for those metrics. We see these results as a first approximation of base rates for the code size and growth properties. Finally, we have identified several next steps for refining and extending our results.

\bibliographystyle{elsarticle-num}
\bibliography{papers-library}

\appendix
\section{Details on SVN configuration validation}
\label{Appendix:svn-details}

The following (case insensitive) regular expressions were used to identify properly submitted SVN code directories. If none of the expressions can be matched on an SVN URL supplied for the project, it is very likely all project branches and tags are included in Ohloh's analysis, and the project should thus be excluded from the data set.\\
\begin{center}
\begin{tabular}{l|l}
\verb!.*/trunk/?! & \verb!.*/head/?! \\
\verb!.*/sandbox/?! & \verb!.*/site/?! \\
\verb!.*/branches/\w+! & \verb!.*/tags/\w+!
\end{tabular}
\end{center}

\section{Boxplots for the metrics}
\label{Appendix:detailed-results}
\begin{figure}[h]
\begin{center}
\includegraphics[width=0.9\textwidth]{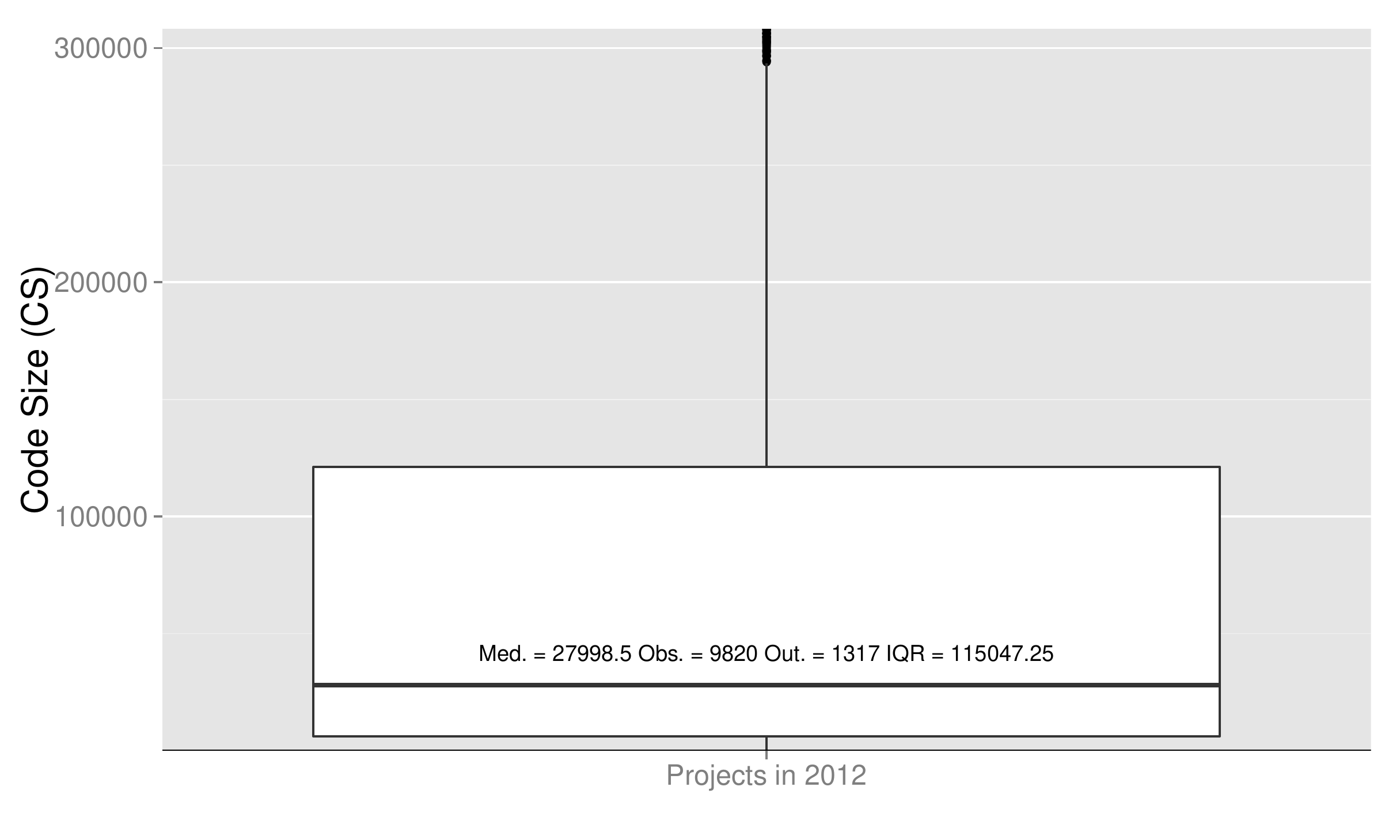}
\caption{Boxplot of the Code Size (CS) metric for projects in 2012. The boxplot is zoomed in such that the whiskers of the plot fill most of the vertical axis. Not all outliers are therefore visible.}
\label{Figure:boxplot-codesize-2012}
\end{center}
\end{figure}

\begin{figure}[h]
\begin{center}
\includegraphics[width=0.9\textwidth]{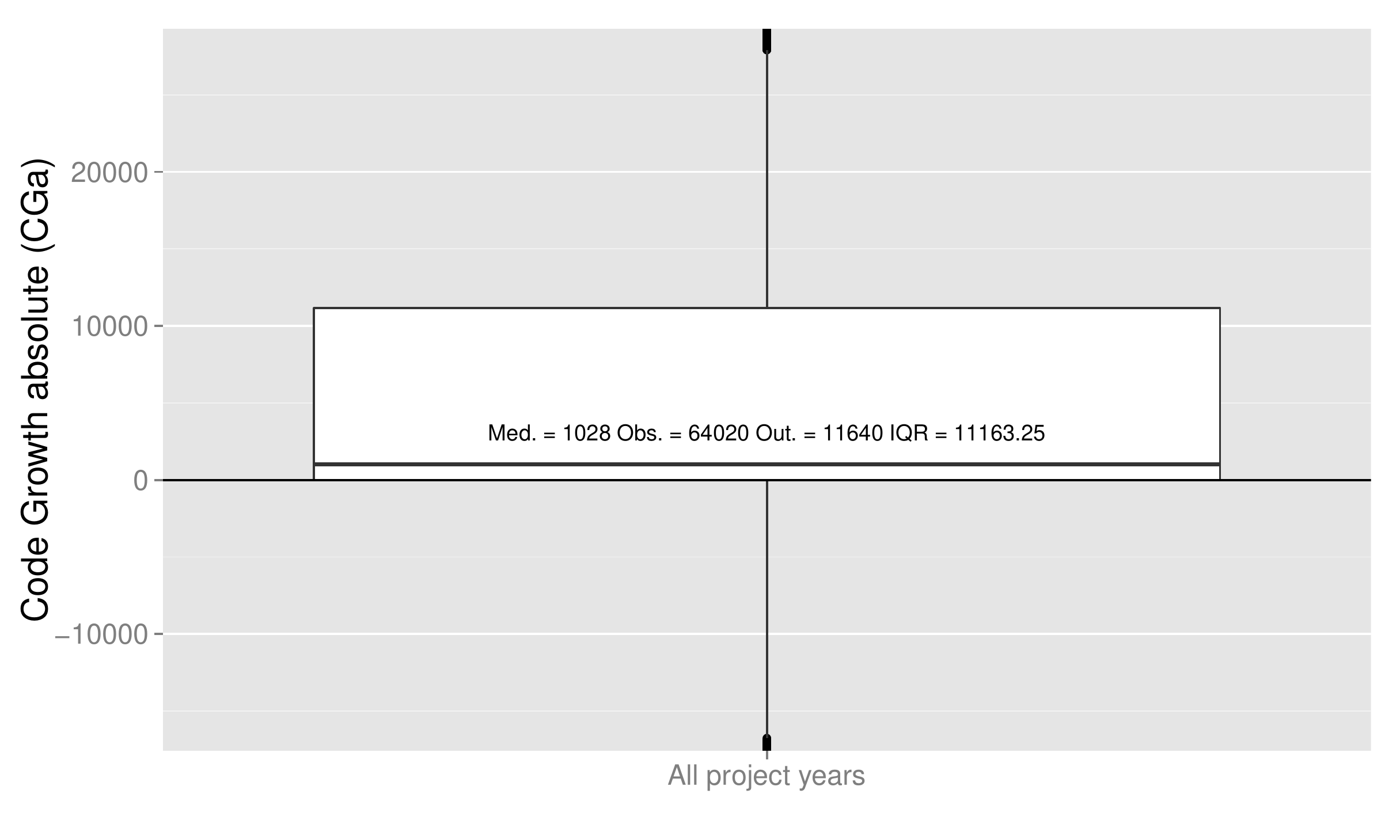}
\caption{Boxplot of the Absolute Code Growth (CGa) metric for all project years. The boxplot is zoomed in such that the whiskers of the plot fill most of the vertical axis. Not all outliers are therefore visible.} 
\label{Figure:boxplot-codegrowth-absolute}
\end{center}
\end{figure}

\begin{figure}[h]
\begin{center}
\includegraphics[width=0.9\textwidth]{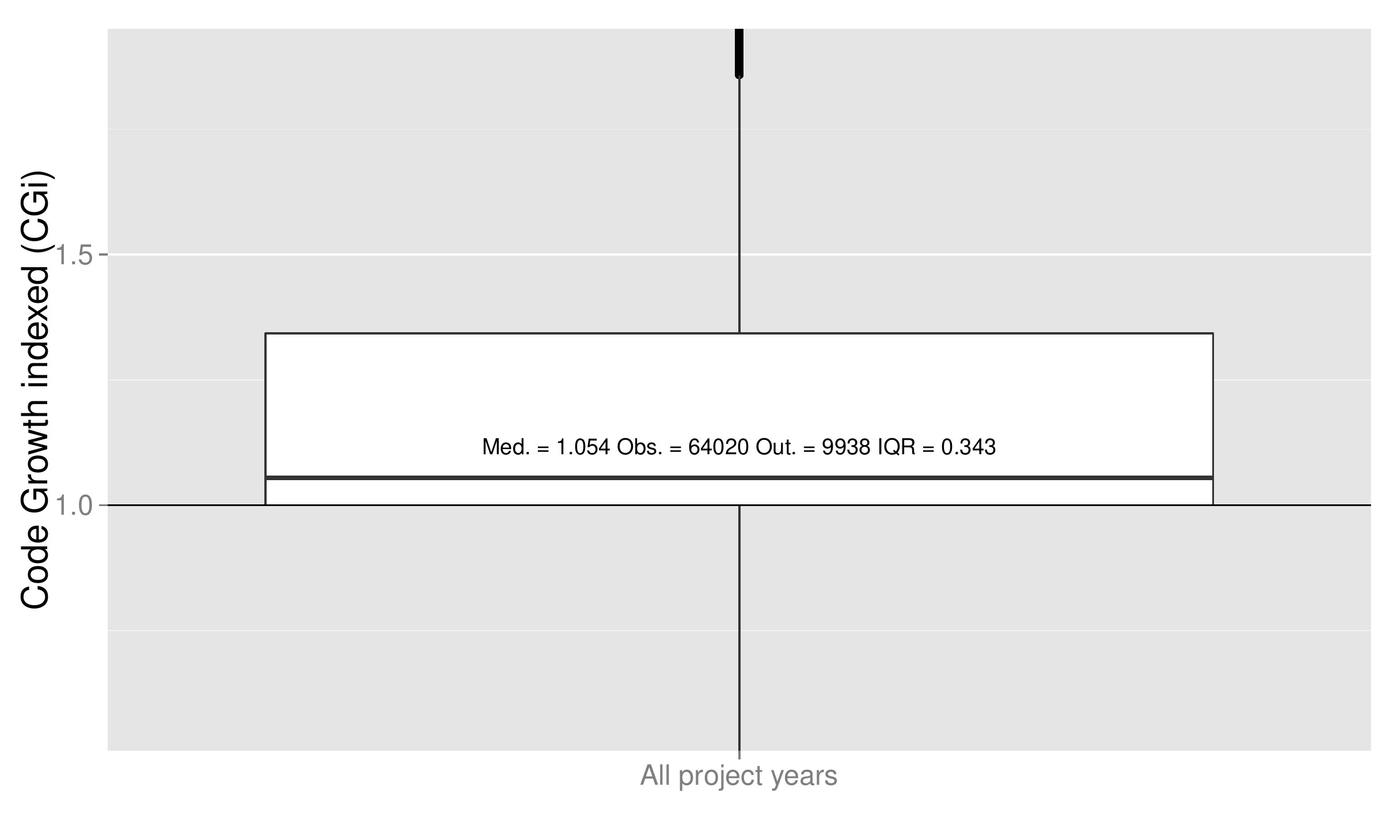}
\caption{Boxplot of the Indexed Code Growth (CGi) metric for all project years. The boxplot is zoomed in such that the whiskers of the plot fill most of the vertical axis. Not all outliers are therefore visible.}
\label{Figure:boxplot-codegrowth-indexed}
\end{center}
\end{figure}


\end{document}